\documentclass[11pt,a4paper,twoside]{article}
\usepackage{epsfig}
\usepackage[latin1]{inputenc}
\usepackage{graphicx,bbm}
\usepackage{amsmath,amsfonts,amssymb}
\usepackage{fleqn} 
\usepackage{mathrsfs}
\usepackage[footnotesize]{caption}
\def\be{\begin{equation}}
\def\ee{\end{equation}}
\def\ba{\begin{array}}
\def\bacc{\begin{array} {cc}}
\def\ea{\end{array}}
\def\bea{\begin{eqnarray}}
\def\eea{\end{eqnarray}}
\def\bd{\begin{displaymath}}
\def\ed{\end{displaymath}}

\def\Box{ {\,\lower 0.9pt\vbox{\hrule\hbox{\vrule height0.2cm \hskip 0.2cm
\vrule height 0.2cm }\hrule}\,}}

\def\<{\left\langle}
\def\>{\right\rangle}

\def\beq{\begin{equation}}
\def\eeq{\end{equation}}
\def\bea{\begin{eqnarray}}
\def\eea{\end{eqnarray}}

\begin{document}


\begin{titlepage}

\vspace*{-15mm}
\begin{flushright}
DCPT-08-54 \\
IPPP-08-27\\
MPP-2008-34\\
SHEP-08-19\\
IC-2008-18\\
NSF-KITP-08-88
\end{flushright}
\vspace*{3mm}

\begin{center}
{
\bf\LARGE
Flavon Inflation
}
\\[8mm]
S.~Antusch$^{\star}$
\footnote{E-mail: \texttt{antusch@mppmu.mpg.de}}, 
S.~F.~King$^{\dagger}$
\footnote{E-mail: \texttt{sfk@hep.phys.soton.ac.uk}},
M.~Malinsk\'{y}$^{\dagger}$
\footnote{E-mail: \texttt{malinsky@hep.phys.soton.ac.uk}},\\
L.~Velasco-Sevilla$^{\clubsuit}$
\footnote{E-mail: \texttt{lvelasco@ictp.it}}
and I.~Zavala$^{\heartsuit}$
\footnote{E-mail: \texttt{ivonne.zavala@durham.ac.uk}}
\\[1mm]

\end{center}
\vspace*{0.50cm}
\centerline{$^{\star}$ \it 
Max-Planck-Institut f\"ur Physik (Werner-Heisenberg-Institut),}
\centerline{\it 
F\"ohringer Ring 6, D-80805 M\"unchen, Germany}
\vspace*{0.1cm}
\centerline{$^\dagger$ \it School of Physics and Astronomy,}
\centerline{\it  
University of Southampton,
Southampton, SO17 1BJ, UK}
\vspace*{0.1cm}
\centerline{$^{\clubsuit}$ \it ICTP, Strada Costiera 11, Trieste 34014 Italy}
\vspace*{0.1cm}
\centerline{$^\heartsuit$ \it CPT and IPPP, Durham University, South Road, DH1 3LE, Durham, UK}
\vspace*{0.50cm}

\begin{abstract}

\noindent 
We propose an entirely new class of particle physics models of
inflation based on the phase transition associated with the
spontaneous breaking of family symmetry responsible for the generation
of the effective quark and lepton Yukawa couplings.  We show that the
Higgs fields responsible for the breaking of family symmetry, called
flavons, are natural candidates for the inflaton field in new
inflation, or the waterfall fields in hybrid inflation. This opens up
a rich vein of possibilities for inflation, all linked to the physics of
flavour, with interesting cosmological and phenomenological
implications.  Out of these, we discuss two examples
which realise flavon inflation: a model of new inflation based on the
discrete non-Abelian family symmetry group $A_4$ or $\Delta_{27}$,
and a model of hybrid
inflation embedded in an existing flavour model with a continuous
SU(3) family
symmetry. With the inflation scale and family symmetry breaking scale
below the Grand Unification Theory (GUT) scale, these classes 
of models are free of the monopole (and similar) problems which are often associated with the GUT phase transition.

\end{abstract}

\end{titlepage}

\newpage

\setcounter{footnote}{0}

\section{Introduction}

Although the inflationary paradigm was proposed nearly thirty years ago
to solve the horizon and flatness problems \cite{Guth:1980zm}, it is
only relatively recently that its main predictions of flatness and density perturbations have been firmly demonstrated to be consistent with observation \cite{Spergel:2006hy,wmap5}.
However the precise nature of the inflation mechanism 
remains obscure and several versions of inflation
have been proposed, bearing names such as old inflation, new
inflation, natural inflation, supernatural inflation, 
chaotic inflation, hybrid inflation, hilltop inflation,
and so on \cite{Lyth:1998xn}. Furthermore, 
the relation of any of these mechanisms for inflation 
to particle physics remains unclear, despite 
much effort in this direction \cite{Lyth:1998xn}.
Indeed it is not even clear if the ``inflaton''
(the scalar field responsible for
inflation) resides in the visible sector of the theory
or the hidden sector. 

On the other hand, one of the great problems facing particle physics is the
flavour problem, i.e.\ the origin of the three families
of quarks and leptons and their Yukawa couplings responsible
for their masses and mixings. In the past decade, the flavour problem has 
been enriched by the discovery of neutrino mass and mixing,
leading to an explosion of interest in this area
\cite{King:2007nw}. A common approach is to suppose that the
quarks and leptons are described by some
family symmetry which is spontaneously broken at a high
energy scale by new Higgs fields called ``flavons''\cite{gflsym}.
In particular, the approximately tri-bimaximal
nature of lepton mixing provides a renewed
motivation for the idea that the Yukawa couplings are
marshalled by a spontaneously broken non-Abelian family symmetry
which spans all three families, for example $SU(3)$
\cite{su3refs}, $SO(3)$ \cite{King:2005bj},
or one of their discrete subgroups 
such as $\Delta_{27}$ or $A_4$ \cite{deMedeirosVarzielas:2005qg}.
Furthermore such family symmetries provide a possible solution to the
supersymmetric (SUSY) flavour and CP problems \cite{Ross:2002mr}.

In this paper we suggest that the phase transition 
associated with the spontaneous breaking of family symmetry
is also responsible for cosmological inflation, a possibility 
we refer to as {\em flavon inflation}.
We emphasise that flavon inflation does not represent 
a new mechanism for inflation, but rather a whole class
of inflation models associated with the 
spontaneous breaking of family symmetry.
For example, the flavons themselves are natural candidates for
inflaton fields in new inflation. 
Since most of the family symmetry models rely on SUSY,
we shall work in the framework of SUSY inflation, with supergravity (SUGRA)
effects also taken into account. In family symmetry models there may
also be other fields associated with the vacuum alignment of the
flavons, often called ``driving'' superfields, and these can alternatively
be considered as candidates for the inflaton, with the
flavons being identified as the ``waterfall fields'' of SUSY hybrid
inflation.\footnote{We also refer to this possibility as flavon inflation,
since the flavon is involved in inflationary dynamics.}

As we shall show, flavon inflation is exceptionally well
suited for driving cosmological inflation.
In new and hybrid inflation, the  inflationary scale is well known
to lie below the GUT scale, causing some tension in
models based on GUT symmetry breaking. 
One advantage of flavon inflation is that the breaking of the family symmetry and hence inflation, can occur below the GUT scale.
Of course an earlier stage of inflation may 
have also occurred at the GUT scale, but it is the lowest scale of inflation
that is the relevant one for determining the density perturbations.
Another advantage of flavon inflation is that 
in inflationary models associated with the breaking of a GUT symmetry
are often plagued by the presence of magnetic monopoles which 
tend to overclose the Universe. In the case of flavon inflation,
since the family symmetry is completely broken\footnote{\label{discrete} We note that if 
at the lowest order in the effective operator expansion some discrete 
symmetries remain, these are broken by higher order effects. 
Possibly created domain wall networks from such remaining discrete
symmetries are therefore in general unproblematic, because they are 
effectively blown away by the pressure generated by these 
higher-dimensional operators.}, no monopoles
result and therefore the monopole problem is absent, and in addition
any unwanted relics associated with the GUT scale breaking are
inflated away.
 
In Section 2 we discuss in general terms how the
idea of flavon inflation opens up a rich vein of possible inflationary scenarios, 
all linked to the physics of flavour, and discuss some of
their interesting cosmological and phenomenological implications. 
In Section 2.1 we briefly review the motivation for family symmetry
and flavons.
In Section 2.2 and 2.3 we consider two concrete examples of
flavon inflation using flavour models. In the first example
we show how new inflation can arise with the flavons in fundamental 
representations of an $A_4$ family symmetry group playing the role of 
inflatons. In the second example, we show how the driving superfields
responsible for the vacuum alignment of the flavons can play the
role of inflatons, with the flavons being the waterfall fields of 
hybrid inflation. In Section 3 we discuss possible implications of
flavon inflation for cosmology and particle physics. 
Section 4 concludes the paper.

 \section{Family symmetry breaking and inflation}\label{Sec:FlavourInflation} 
\subsection{Family Symmetry and Flavons}
One of the greatest mysteries facing modern particle physics is
that of the origin of quark and lepton masses and mixings.
In the Standard Model (SM) they arise from Yukawa matrices 
and (in the see-saw extended SM) right-handed neutrino Majorana
masses. In order to understand the origin of fermion masses
and mixing, a common approach is to assume that the SM is extended
by some horizontal family symmetry $G_F$, which may be continuous
or discrete, and gauged or global. It must be broken completely,
apart from possibly remaining discrete symmetries,
at some high energy scale in order to be phenomenologically
consistent, and such a symmetry breaking requires the introduction
of new Higgs fields called flavons, $\phi$, whose vacuum 
expectation values (vevs) break the family symmetry $\<\phi \>\neq 0$.

The Yukawa couplings are forbidden
by the family symmetry $G_F$, but once it is broken, effective Yukawa
couplings may be generated by non-renormalizable operators
involving powers of flavon fields, for example 
$(\phi /M_c)^n\psi \psi^c H$ leading to an effective Yukawa
coupling $\varepsilon^n\psi \psi^c H$ 
where $\varepsilon=\<\phi \>/M_c$ and  $\psi, \psi^c$ are SM fermion fields,
 $H$ is a SM Higgs field, and $M$ is some high energy mass scale associated with
the exchange of massive particles called messengers.
Phenomenology requires typically $\varepsilon \sim 0.1$.

If in addition to the family symmetry, the SM gauge group
is unified into some GUT gauge group 
$G_\mathrm{GUT}$ (for example $SU(5)$, $SO(10)$, etc.)
then the high energy theory has the symmetry structure
$G_F\times G_\mathrm{GUT}$. In such frameworks, the theory
has additional constraints arising from the fact that
the messenger sector must not spoil unification.
This implies that either 
the messenger sector scale $M_c$ has to be very
close to the GUT scale $M_\mathrm{GUT}$ (thus pushing also the family
symmetry breakdown close to the GUT scale) or the messengers
must come in complete GUT representations,
leading to consequences for low energy phenomenology.
Assuming that the flavon sector is responsible for
inflation provides additional information on the scale
of family symmetry breaking, as we now discuss
in the framework of two examples.

\subsection{Example 1: Flavon(s) as inflaton(s)}

The first example we discuss
is where the flavon plays the role of the inflaton in a
new inflation model,  similar to the one
discussed in \cite{Senoguz:2004ky}. However, we make use of the fact
that when the inflatons are representations of family symmetry groups
instead of GUT groups, new possibilities for inflation models arise.
To be more explicit, in the considered class of inflation models, in
addition to the invariant combination of fields $(\bar\phi
\phi)^n/M_*^{2n-2}$ (with $\phi$ and $\bar\phi$ being two fields in
conjugate representations) studied in \cite{Senoguz:2004ky} we can
now write any combination of family-symmetry-invariant fields. For
example with  a non-Abelian discrete family symmetry $A_4$
or $\Delta_{27}$,
superpotentials of the form
\begin{eqnarray}\label{Eq:InvHybrid}
W = \kappa S \left[ \frac{(\phi_1 \phi_2 \phi_3)^n}{M_*^{3n-2}} - \mu^2  \right]
\end{eqnarray} 
with $n \ge 1$ can appear, since for fields $\phi =
(\phi_1,\phi_2,\phi_3)$, $\psi = (\psi_1,\psi_2,\psi_3)$ and $\chi =
(\chi_1,\chi_2,\chi_3)$ each in the fundamental triplet $\underline{3}$
representation of $A_4$ or $\Delta_{27}$,
the combination $\{\phi_1 \psi_2 \chi_3$ + permutations$\}$ forms an
invariant. Thus the above combination $\phi_1 \phi_2 \phi_3$
corresponds to a particular $A_4$ 
or $\Delta_{27}$
invariant $\phi^3=\phi_1 \phi_2 \phi_3$.
Without loss of generality the Yukawa coupling $\kappa$ can be set
equal to unity as in \cite{Senoguz:2004ky}. At the global minimum
of the potential the $\phi_i$ components get vevs
of order $M=M_*\left(\frac{\mu}{M_*}\right)^{2/3n}$.

\begin{figure}[t]
\centering
$\ensuremath{\vcenter{\hbox{\includegraphics[scale=0.885]{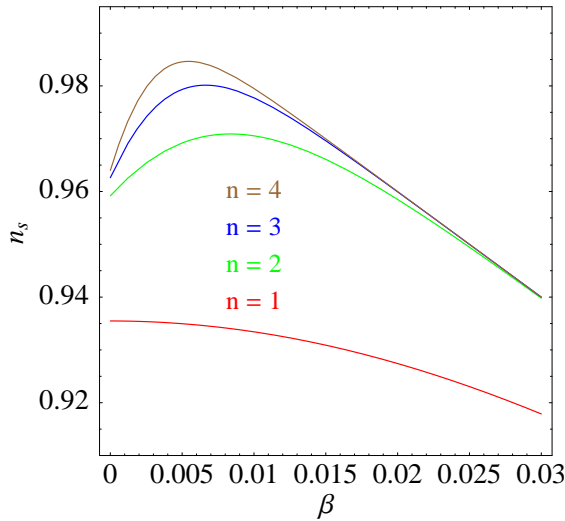}}}}$ \\[2mm] 
 \caption{\label{fig:nsresults} 
Predictions for the spectral index $n_s$ as a function of $\beta$ for $N=60$.  
For comparison, the results for the invariant combination of fields $(\bar\phi \phi)^n/M^{2n-2}$ for different $n$ can be found in figure 1 of \cite{Senoguz:2004ky}. }
\end{figure}

In the following we analyse the phenomenological predictions of this
class of models.  We assume a K\"ahler potential of the non-minimal
form (analogous to \cite{Senoguz:2004ky})
\begin{eqnarray}
 K &=& |S|^2 + |\phi|^2 +  \kappa_2 \frac{|S|^2 |\phi|^2}{M_P^2} + 
 \kappa_1\frac{|\phi|^4}{4 M_{Pl}^2} + \kappa_3\frac{|S|^4}{4 M_{Pl}^2} + ... 
\label{eq:inf_flav_A4}
\end{eqnarray}
and study the supergravity F-term scalar potential\footnote{Since
 $A_4$ is a discrete symmetry, there are no D-term contributions                               
to the potential. },    which is given by
\beq\label{sugrapot} 
V= e^{K/M_{Pl}^2} \, \left[ K^{i\bar
\jmath}D_{i}W D_{\bar\jmath}\overline{W} - 3\frac{|W|^2}{M_{Pl}^2}
\right], 
\eeq 
where $D_iW = \partial_i W +\frac{W}{M_{Pl}^2} \,
\partial_i K$ and $K^{i\bar \jmath} = (\partial_{i}\partial_{\bar
\jmath} K)^{-1}$. In order to study this potential, we assume $S\ll
M_{Pl}$ and $\phi_i \ll M \ll M_{Pl}$, where $M_{Pl} $ is the reduced Planck mass $M_{Pl} = (8\pi G_N)^{-1/2}\sim 10^{18}$ GeV.  During inflation, we consider
the case in which the driving field $S$ acquires a large mass and
therefore goes rapidly to a zero field value
(which can be achieved by choosing
$\kappa_3 < -1/3$ such that it is heavier than the Hubble scale
\cite{Senoguz:2004ky}).  Furthermore, we focus on the situation in
which the component fields $\phi_i$ start moving from close to zero
(the local maximum of the potential) and roll slowly towards the true
minimum of the potential where $\phi_i \sim M$ (vacuum dominated
inflation \cite{Copeland:1994vg}).  It is possible to show that a
generic inflationary trajectory occurs when all components, $\phi_i$, of the
triplet field are equal. 
Therefore, we concentrate on this trajectory in what follows\footnote{When this is not the case, we have in general a multifield flavon inflationary
scenario. This can arise in family symmetry models as considered for
instance in \cite{deMedeirosVarzielas:2005ax}, as part of a multistage
inflationary model. A detailed analysis of this situation is left for
a future publication. }.  Defining the real field components as
$|\phi_i| \equiv \varphi /\sqrt{2}$ and $\beta = (\kappa_2 -1)$,
$\lambda = (\beta(\beta+1)+1/2+\kappa_1/12)$ and $\gamma =
2/(6)^{3n/2}\lesssim 0.14$, we obtain 
 the potential during inflation \cite{ref:potpwrd}: 
 \be V \simeq \mu^4 \left[ 1 -
\frac{\beta}{2} \frac{\varphi^2}{ M_{Pl}^2} +
\frac{\lambda}{4}\frac{\varphi^4}{ M_{Pl}^4} - \gamma
\frac{\varphi^{3n}}{M^{3n}} + \cdots\right].  \ee 
In the following, we consider the situation where
$|\gamma \frac{\varphi^{3n}}{M^{3n}}| \gg
|\frac{\lambda}{4}\frac{\varphi^4}{ M_{Pl}^4}|$. Thus the quartic term
$\varphi^4 /M_{Pl}^4 $ can be neglected{\footnote{When $|\gamma
    \frac{\varphi^{3n}}{M^{3n}}| \ll
    |\frac{\lambda}{4}\frac{\varphi^4}{ M_{Pl}^4}|$  it turns out that
    $M\sim M_{Pl}$, $\mu\sim 10^{15}$ GeV and so $M_*(> M)\sim
    M_{Pl}$, which is not very useful information about the family
    symmetry breaking scale. Inflation can be obtained but it requires
    some fine tuning of the parameter $\lambda$ \cite{flavinfl_ds}.}} and we find that the spectral index  $n_s$, can be expressed in terms of the parameters of the potential and the number of e-folds $N$ as:\footnote{In standard slow roll inflation $n_s -1 =2\eta -6\epsilon$, where $\eta,\, \epsilon$ are the standard slow roll
parameters. In the present case we have $\epsilon\ll \eta$.}
\bea n_s &\approx& 1 - 2 \beta \left[ 1+
\frac{(3n-1)(1-\beta)} {[(3n-2)\beta + 1] e^{\beta(3n-2)N} +\beta -1}
\right] \qquad {\rm for \,\,\, } \beta \neq 0 \;,
\\
n_s &\approx& 1 -  \frac{6n-2}{(3n-2)N + (3n-1)}
 \,\,\, \qquad\qquad\qquad\qquad\qquad \;\;{\rm for \,\,\, } \beta = 0 \;.
\eea
The results are illustrated in figure \ref{fig:nsresults}. The predictions for $n_s$ are close to the WMAP 5 year data \cite{wmap5} $n_s = 0.960 \pm 0.014$ for $n\ge 2$ and $\beta \lesssim 0.03$. In all cases we have taken $N = 60$. 

\begin{figure}
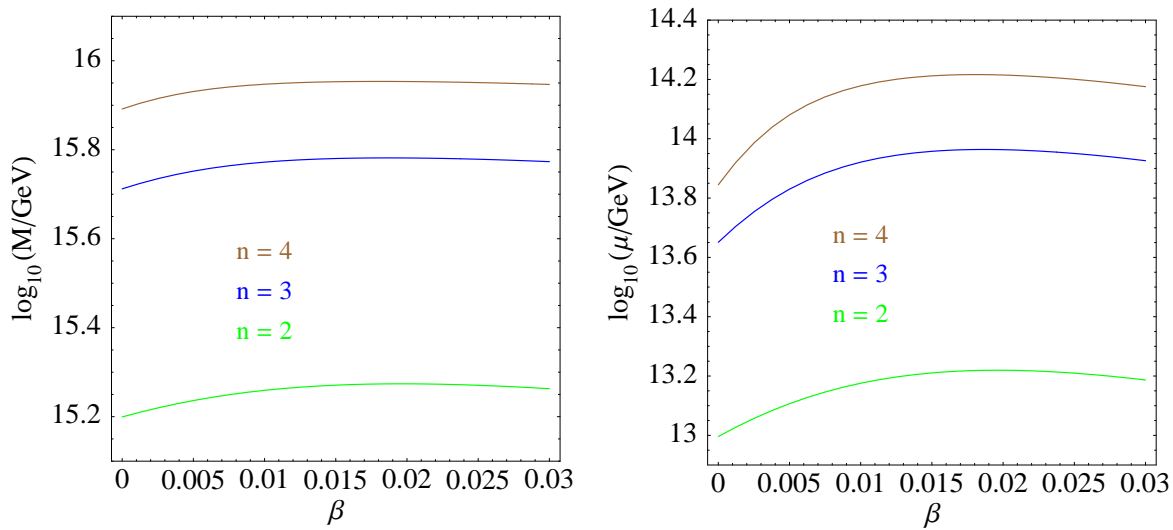

\centering
$\ensuremath{\vcenter{\hbox{\includegraphics[scale=0.885]{M.eps}}}}$ \;\;
$\ensuremath{\vcenter{\hbox{\includegraphics[scale=0.885]{mu.eps}}}}$\\[2mm]
\caption{\label{fig:MandMuresults} 
Predictions for the family breaking scale $M$ (left plot) and for the inflation scale $\mu$ (right plot) as a function of $\beta$ for $N=60$ and $M_* = M_\mathrm{GUT}$ ($M_{GUT} = 2\times 10^{16}$GeV).}
\end{figure}

The scale $M$, which governs the size of the flavon vev $\phi$, and the inflation scale $\mu$ are determined by the temperature fluctuations $\delta T/T$ of the CMB (assuming that inflation and $\delta T$ originate from $\phi$) for a given generation scale $M_*$ of the effective operator in Eq.~(\ref{Eq:InvHybrid}). 
Specifically, we can relate the scale $M_*$ to $M$ and $\mu$ via the amplitude of the density perturbation when it re-enters the horizon, 
\begin{eqnarray}
\delta_H = \frac{1}{5\pi\sqrt{3}} \frac{V^{3/2}}{M_{Pl}^3|V'|} = 1.91 \times 10^{-5}\;.
\end{eqnarray} 
If we write this equation explicitly in terms of $\mu^2$ and $M$ and  relate it to $M_*$, as defined below Eq.~(\ref{Eq:InvHybrid}) we get:
\bea\label{masses}
M^{\frac{9n(n-1)}{3n-2}}\!\! & = & \!\! 
M_*^{3n-2}M_{Pl}^{\frac{3n-4}{3n-2}}\,5\pi\sqrt{3}\,\delta_H
\left[ \frac{\beta(1-\beta)}{3n\gamma\left[((3n-2)\beta +1)e^{(3n-2)N}+\beta-1\right]}   \right]^{\frac{1}{3n-2}}   \nonumber \\
&&\hskip 1cm  \cdot \left[ \beta+ 
      \frac{\beta(1-\beta)}{((3n-2)\beta +1)e^{(3n-2)N}+\beta-1} \right],
 \qquad {\rm for \,\,\, } \beta \neq 0,  \\
M^{\frac{9n(n-1)}{3n-2}}\!\! &= & \!\!
M_*^{3n-2}M_{Pl}^{\frac{3n-4}{3n-2}}\,5\pi\sqrt{3}\,\delta_H
\left[\frac{1}{3 n \gamma \left[(3n-2)N +3n -1\right]}\right]^{\frac{1}{3n-2}} \nonumber \\
&&\hskip 3.5cm  \cdot \left[\frac{1}{((3n-2)N+3n-1} \right],   
 \qquad {\rm for \,\,\, } \beta = 0  \;.    
\eea
With $M_*$ around the GUT scale ($M_{GUT} = 2\times 10^{16}$GeV), we obtain (again $N=60$)
\bea
  M &\approx& (10^{15},10^{16})  \;\mbox{GeV}, \quad n=2,3,4 \;, \\
\mu &\approx& (10^{13},10^{14})  \;\mbox{GeV}, \quad n=2,3,4 \;,  
\eea
for all considered values of $\beta$.
The numerical results are shown in figure \ref{fig:MandMuresults}. We note that the choice $M_* = M_\mathrm{GUT}$ is only an example. In principle $M_*$ can be much lower and thus also $M$ would be much lower (for example an intermediate scale $M_*  \approx (10^{11},10^{13}) \;\mbox{GeV}$ can give $M\approx (10^{11},10^{13})\;\mbox{GeV} $, $\mu \approx (10^{10},10^{12})\;\mbox{GeV}$). 

For $n=1$ and $N\sim 60$ ($\beta$ in the relevant range), $M_*$ is not longer a free parameter and in fact it is determined to be around $10^{24}$ GeV. 
In this case we are free to choose $\mu$ and thus we could in principle have low scale inflation under this condition. Also, in order to have $M < M_\mathrm{GUT}$, $\mu$ would have to be below about $10^{10} - 10^{11}$ GeV. However, since $M_*$ is found to be larger than the Planck Scale, it cannot be regarded as a fundamental generation scale of the effective operator but itself has to emerge as an effective scale. 

When (at least part of) the family symmetry breaking takes place below $M_\mathrm{GUT}$, this has interesting phenomenological consequences as we discuss in Sec.~\ref{Sec:Pheno}. 

\subsection{Example 2: Driving superfield(s) as inflaton}  
In supersymmetric theories the superpotentials which determine the
flavons' vevs contain another class of fields in addition to the
flavons, the so-called {\em driving superfields}. The driving
superfields are singlets under the family symmetry, in contrast to
the flavons.

As an example of how inflation may be realised from the driving
superfields, we consider a vacuum alignment potential as 
 studied in the SU(3) family symmetry model of
\cite{deMedeirosVarzielas:2005ax}.  We assume the situation that 
$\<\phi_{23} \> \propto (0,1,1)^T$ and $\<\Sigma \>= \mbox{diag}(a,a,-2a)$ are
already at their minima, and that the relevant part of the
superpotential which governs the final step of family symmetry
breaking is given by \cite{deMedeirosVarzielas:2005ax}
\begin{eqnarray}
W = \kappa S (\bar\phi_{123} \phi_{123} - M^2) + \kappa' Y_{123} \bar\phi_{23} \phi_{123}
+ \kappa'' Z_{123} \bar\phi_{123} \Sigma \phi_{123} + ... \;.
\label{mvr_w23}
\end{eqnarray}
$S$ is the driving superfield for the flavon $\phi_{123}$, i.e.\ the contribution to the scalar potential from $|F_S|^2$ governs the vev $\< \phi_{123} \>$.   
In addition we assume a non-minimal K\"ahler potential of the form 
\bea\label{kahler1}
K&=&
|S|^2 + |\phi_{123}|^2 + |\bar\phi_{123}|^2 + |Y_{123}|^2 + |\bar\phi_{23}|^2 + |\phi_{23}|^2 + |Z_{123}|^2 + |\Sigma|^2 \nonumber \\
&&
+ \kappa_S\frac{|S|^4}{4 M_{Pl}^2}
+ \kappa_{SZ}\frac{|S|^2 |Z_{123}|^2}{4 M_{Pl}^2}
+ ...
\;.
\eea
Although the theory is rather complicated we emphasise that 
this is taken from the existing family symmetry literature. For the purposes of inflation, we are interested
in the epoch where the fields with larger vevs 
$Y_{123},\,\phi_{123},\,\bar\phi_{123}$ do not evolve, and inflation is 
provided by the fields with smaller vevs. 
We note that due to the vevs of $\phi_{23}$ and $\Sigma$, SU(3) is already broken. In order to proceed, we analyse the supergravity scalar potential  (\ref{sugrapot}) focusing on the D-flat directions which are potentially promising for inflation. Setting $Y_{123}=\phi_{123}=\bar\phi_{123}=0$ since the fields obtain large masses from the superpotential, the tree level scalar potential takes the simple form (expanded in powers of fields over $M_{Pl}$)\footnote{Here we show only the relevant terms for inflation. However one should keep in mind that quartic terms are present such that the fields are evolving from large values to small ones. The details of this model have been presented in 
\cite{BasteroGil:2006cm}. } 
\bea
V =  \kappa^2 M^4\,   \left[ 1 -  \gamma \frac{\xi^2}{2M_{Pl}^2}   
- 2\kappa_{S} \frac{\sigma^2}{2 M_{Pl}^2} + ...\right],
      \label{sugrapoteff1}
\eea
where we have defined $|S| = \sigma/\sqrt{2}$, $|Z_{123}| = \xi/\sqrt{2}$ and $\gamma = \kappa_{SZ}-1$. 
From this expression, we see that if the two coefficients in front of the mass terms for $\sigma$ and/or $\xi$ are sufficiently small both/one of them can drive inflation. 

If we assume that $\sigma$ acts as the inflaton (choosing for instance $\gamma < -1/3$ such that the mass of $\xi$ exceeds the Hubble scale) and taking into account loop corrections to the potential, it has been shown in \cite{lr,BasteroGil:2006cm} that for $\kappa_{S} \approx (0.005 - 0.01)$ and $\kappa \approx (0.001 - 0.05)$, a spectral index consistent with WMAP 5 year data \cite{wmap5}, $n_s = 0.96 \pm 0.014$, is obtained. 
Finally, the scale $M$ of family symmetry breaking along the $\< \phi_{123} \>$-direction is determined 
from the temperature fluctuations $\delta T/T$ of the CMB to be
\begin{eqnarray}
M \approx 10^{15} \;\mbox{GeV}\;,
\end{eqnarray}
about an order of magnitude below the GUT scale. After having analysed
two example scenarios, let us now turn to a general qualitative 
discussion of possible consequences of flavon inflation.

\section{Discussion and Implications of Flavon Inflation}\label{Sec:Pheno} 
The connection between family symmetry breaking and inflation has
several implications for theories of inflation as well as for
theories of flavour.  In this section we discuss some of the
cosmological and particle physics consequences.

Many important implications are related to the fact that the
scale of family symmetry breaking (which is connected to the scale of
inflation) is determined by the temperature fluctuations of the CMB, 
i.e.\ inflation predicts the 
scale of family symmetry breaking.
In the two examples presented in
sections 2.2 and 2.3, we have found that (at least the relevant part
of the) family symmetry breaking takes place at about $10^{15}$
to $10^{16}$ GeV, that is, below the GUT scale. Another intriguing
possibility would be flavon inflation at TeV energies, such that
both, the flavour sector and the inflationary dynamics, might be
observable at the LHC. 

One attractive feature of having inflation after a possible GUT phase
transition is that unwanted relics from the GUT phase transition, such
as monopoles, are diluted. Furthermore, after spontaneous family
symmetry breaking the symmetry is commonly completely
broken, which means that, in particular, no continuous symmetry
remains. Possibly created domain wall networks from remaining discrete
symmetries are in general unproblematic, because they are 
effectively blown away by the pressure generated by higher-dimensional 
operators which break the discrete symmetries (c.f.\ comment in footnote 2).

The fact that in some cases (as discussed in the text) inflation can predict the scale of family symmetry breaking
to be below the GUT scale, can give rise to other additional consequences. For example,
the renormalisation group (RG) evolution of the SM quantities from low
energies to the GUT scale is modified by the new physics at
intermediate energies.  Thus, the predicted GUT scale ratios of
Yukawa couplings from low energy data will be modified due to the
intermediate family symmetry breaking scale, which would affect, for instance, the
possibility of third family Yukawa unification.  Furthermore, the
predictions for the fermion masses and mixings emerge at the family
symmetry breaking
scale. The knowledge of this scale is important for precision
tests of these predictions. In particular, the renormalisation group
running between this scale and low energy has to be taken into
account.

The idea of flavon inflation yields new possibilities
for inflation model building, and in general predicts a rich variety of
possible inflationary trajectories for single field and also
multi-field inflation. In the example studied in section
2.2 we found that if the flavon(s) in fundamental representations of
an $A_4$ or $\Delta_{27}$
family symmetry act as the inflation(s), there are new
invariant field combinations in the (super)potential which have not
been considered for inflation model building so far. In the example in
section 2.3 we have seen that in addition to types of models similar
to standard SUSY hybrid inflation, it is also generic (depending on
the parameters of the K\"ahler potential) that the scalar components
of more than one driving superfield participates in inflation. In
addition, it is typical that family symmetry breaking proceeds in
several steps, which would mean that before the observable inflation
there could have been earlier stages of inflation.

Besides  the rich structure of the potentials during inflation,
there is also  an interesting dynamics of the flavon fields
after inflation. The potentials are usually  such
that not only the moduli of the vevs of the flavons are determined,
but also that they point into specific directions in flavour space.
When the flavons are moving towards their true 
values after inflation, the dynamics of the field evolution often has
a much larger complexity and diversity than conventional inflation
models. This is due to the fact that typically several flavon components are
moving, and that the potentials have special shapes in order to force
the flavon vevs to point in the desired directions in flavour space in
the true minimum.  

This can have consequences for the density perturbations (non adiabaticities, non-Gaussianities, etc.) as well as for baryogenesis during preheating and during (and after) reheating.
Non-thermal leptogenesis, for example, would be
connected to the physics of family symmetry breaking and new
possibilities for generating the baryon asymmetry will appear. 
Since flavons can play the role of  either the inflaton or the waterfall fields, 
their decays into leptons will be determined by couplings which
are associated with some particular flavour model.
Thus successful non-thermal leptogenesis following flavon inflation
can  provide  further constraints on flavon inflation models,
leading to possible bounds on right-handed neutrino masses
and so on.
There could also be further constraints coming from the proton decay in the unified flavon inflation models. For instance, as the scale of family symmetry breaking $M$ approaches the GUT scale $M_{\mathrm{GUT}}$, the efficiency of monopole dilution is expected to fall. If, on the other hand, $M$ is significantly below
$M_{\mathrm{GUT}}$, the monopoles will be inflated away, but the flavour messenger sector will affect gauge coupling unification, with implications for proton decay. It is also interesting to note that when global family symmetries break, there could be pseudo-Goldstone bosons appearing with interesting phenomenology.

\section{Conclusion}\label{Conclusion} 
In conclusion, we have shown that existing models based on a spontaneously
broken family symmetry, proposed to resolve the flavour problem are naturally linked to cosmology. They
introduce new and promising possibilities for cosmological inflation,
which we have referred to generically as flavon inflation.
In flavon inflation,
the inflaton can be identified with either one of the flavon fields 
introduced to break the family symmetry, or with one of the driving
fields used to align the flavon vevs. In either case the 
scales of inflation and family symmetry breaking result to be typically below the GUT scale in the presented examples.
Since the family symmetry is broken completely (c.f.\ comment in footnote 2) 
this provides a natural resolution of GUT scale cosmological relic problems,
without introducing further relics.
Moreover flavon inflation has  
a large number of interesting consequences for particle
physics as well as for cosmology, which we have only briefly
touched on here but which are worth exploring in more detail in
future studies \cite{flavinfl_ds}.

\section*{Acknowledgements}

L. V-S. and I. Z. would like to thank the KITP Santa Barbara for partial support from NSF under grant No. PHY05-51164, and hospitality during the workshop "String Phenomenology",  where part of this work was discussed. I. Z. thanks also Perimeter Institute for partial support and hospitality while part of this work was done. I. Z. is supported by a STFC Postdoctoral Fellowship. S.A. acknowledges partial support by The Cluster of Excellence for Fundamental Physics ``Origin and Structure of  the Universe'' (Garching and Munich).  S.F.K. and M.M. acknowledge partial support from the following grants: PPARC Rolling Grant PPA/G/S/2003/00096; EU Network MRTN-CT-2004-503369; EU ILIAS RI I3-CT-2004-506222; NATO grant PST.CLG.980066. Finally we thank L. Boubekeur and J. Kersten for helpful comments and the organisers of the CERN workshop 'Flavour in the Era of the LHC' and those of Planck'07 in Warsaw, where this collaboration was started.

\providecommand{\bysame}{\leavevmode\hbox to3em{\hrulefill}\thinspace}

\end{document}